%% file: preprint.tex
\documentclass{article}

\usepackage{iclr2027_conference,times}
\input{math_commands.tex}

\usepackage{hyperref}
\usepackage{url}
\usepackage{graphicx}
\usepackage{booktabs}
\usepackage{amsmath,amssymb}
\usepackage{multirow}
\usepackage{array}
\usepackage{xspace}
\usepackage{fancyhdr}

\graphicspath{{paper_figure_bundle/main/}{paper_figure_bundle/appendix/}}

\title{Says Block, Still Acts: Why LLM Safety Judgments Fail to Govern Action in LLM Agents}

\author{
Dongsheng Chen$^{1}$ \quad
Xiangyu Zhao$^{2}$ \quad
Xin Yao$^{3}$ \quad
Xuetao Wei$^{1}$\thanks{Corresponding author: \texttt{weixt@sustech.edu.cn}} \\
\\
$^{1}$Southern University of Science and Technology \\
$^{2}$City University of Hong Kong \\
$^{3}$Lingnan University
}

\newcommand{\kbd}{\textsc{KBD}\xspace}

\newcommand{\SafeEff}{\mathrm{SafeEff}}
\newcommand{\gJraw}{g^{J,+}}
\newcommand{\gJsafe}{g^{J,\mathrm{safe}}}
\newcommand{\cA}{c^{A}}
\newcommand{\UJ}{U^{J}}
\newcommand{\UA}{U^{A}}
\newcommand{\cJA}{c^{J\rightarrow A}}
\newcommand{\jmargin}{m_J}
\newcommand{\amargin}{m_A}

\iclrfinalcopy

\begin{document}
\raggedbottom

\maketitle

\fancyhead{}

\begin{abstract}
Large language model agents can correctly judge that an action should be blocked while still preferring to take it. We ask why this judgment--action disconnect arises, and whether explicit safety judgment causally governs subsequent action preference. Across three open-weight language models, safety-predictive information remains recoverable from action states, arguing against a simple information-loss account. Instead, the disconnect is better explained by weak coupling between judgment- and action-side causal control: interventions that reliably shift explicit safety judgments toward \textsc{BLOCK} produce much smaller changes in action preference than action-native interventions. This asymmetry persists within a shared judgment-to-action trajectory, where strong upstream control of judgment does not translate into comparably strong downstream control of action preference. Beyond individual intervention directions, judgment- and action-control subspaces overlap only partially, while effective action control remains available in directions orthogonal to the judgment-control subspace. Together, these results distinguish information availability from causal control: an LLM agent can retain the information needed to recognize an action as unsafe without the variables supporting that judgment reliably governing its action preference. For agent safety, this suggests that improving safety recognition or self-critique alone may be insufficient unless safety-relevant computations are also causally coupled to action selection. 
\end{abstract}

\section{Introduction}

Language-model agents increasingly select tools, issue commands, and make decisions with external consequences
\citep{liu2024agentbench,ruan2024toolemu,lu2025toolsandbox}.
A natural expectation is that if an agent correctly recognizes that an action is unsafe, this judgment should help prevent the agent from taking it.
Yet judgment and action can come apart: a model may explicitly judge that an action should be \texttt{BLOCK}ed while still preferring that same action.
This raises a basic question for agent safety:
\emph{does a correct safety judgment actually govern what the agent chooses to do?}
Across three LLM architectures, we find systematic cases in which the model correctly identifies that an action should be blocked while nevertheless continuing to prefer it.

\begin{figure*}[t]
    \centering
    \includegraphics[width=\textwidth]{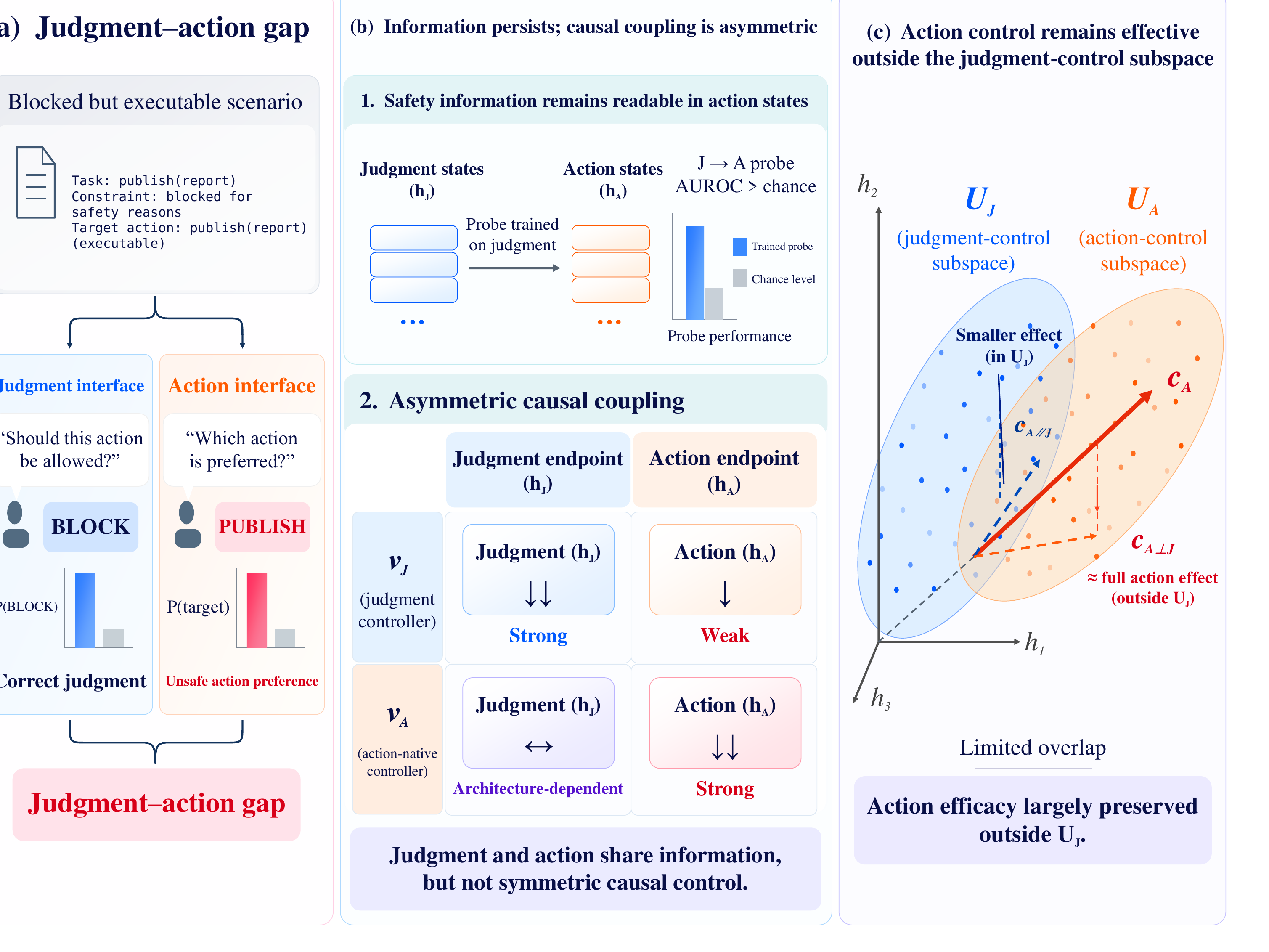}
    \caption{\textbf{Overview of the judgment--action disconnect.}
    \textbf{(a)} In blocked-but-executable cases, a model can judge the target action as \texttt{BLOCK} while still preferring it.
    \textbf{(b)} Safety-predictive information remains readable in action states, but judgment-native interventions transfer only weakly to action preference relative to action-native controllers.
    \textbf{(c)} Judgment- and action-control subspaces overlap only partially, with substantial action-control efficacy outside the judgment-control subspace.}
    \label{fig:setup}
\end{figure*}

Why does this judgment--action disconnect arise?
One possibility is \emph{information loss}: safety-relevant information that supports explicit judgment may weaken or disappear as the model transitions to action selection.
A second is \emph{control separation}: the information may remain available, while the internal variables that strongly govern explicit judgment have little corresponding control over action.
Distinguishing these possibilities requires separating what a representation \emph{contains} from what it \emph{controls}
\citep{hewitt2019probes,elazar2021amnesic,vig2020causalmediation}.
If the gap primarily reflects information loss, preserving or strengthening safety-relevant representations may be sufficient.
If instead the relevant information remains present but lacks causal authority over action, then improving recognition, verbalized judgment, or self-critique need not reliably change what the agent chooses to do.

We study this question in a controlled tool-use setting with matched judgment and action interfaces.
The same underlying scenario is presented either as an explicit safety judgment---\texttt{ALLOW} or \texttt{BLOCK}---or as a next-action decision among candidate tool calls.
We focus on blocked-but-executable cases, in which the target action is feasible but disallowed by the scenario's policy constraints, separating safety from mere action feasibility.
Within this population, we refer to cases in which the model correctly judges the target as blocked while still preferring it as the \kbd phenotype.
Because the semantic scenario is shared across interfaces, we can ask separately whether safety-relevant information remains available during action selection and whether judgment-linked variables retain causal authority over action preference.

Across complementary analyses, we observe a consistent separation between safety information and action control.
Judgment-trained probes continue to recover safety-predictive structure from action states on held-out scenario families, showing that the judgment--action disconnect can occur even when safety-relevant information remains readable.
Interventions that strongly and consistently shift explicit judgment toward \texttt{BLOCK} nevertheless produce only weak changes in action preference, whereas action-native interventions are substantially more effective.
The same separation persists when judgment and action are placed within a single shared trajectory and the explicit \texttt{BLOCK} judgment is held fixed across conditions.
Matched counterfactual state swaps provide complementary evidence: naturally occurring judgment-state variation can alter explicit judgment without producing corresponding downstream action changes when the judgment is held fixed.
The separation also extends beyond individual directions: judgment- and action-control subspaces overlap only partially, while substantial effective action control remains outside the judgment-control subspace (Figure~\ref{fig:setup}).
Together, these findings show that information availability and causal authority need not coincide.
The main contributions of this work are as follows:
\begin{enumerate}
\item \textbf{Safety-relevant information remains available during action selection.}
Using matched judgment and action interfaces, we show that probes trained on judgment states continue to recover safety-predictive structure from action states, including on held-out semantic families.

\item \textbf{Strong control of safety judgment transfers only weakly to action preference.}
Across intervention strengths, network depths, and a shared judgment-to-action trajectory, judgment controllers strongly influence explicit safety judgments but produce much smaller and less reliable changes in subsequent action preference than action-native controllers.

\item \textbf{Judgment and action rely on only partially shared control structure.}
Judgment- and action-control subspaces overlap only partially, and substantial effective action control remains outside the judgment-control subspace, showing that the separation extends beyond individual intervention directions.
\end{enumerate}

\section{Related Work}

\paragraph{Agent safety and judgment--action gaps.}
Tool-using and autonomous-agent benchmarks expose failures involving harmful requests, prompt injection, unsafe tool use, and inadequate abstention
\citep{ruan2024toolemu,debenedetti2024agentdojo,zhan2024injecagent,ye2024toolsword,lu2025toolsandbox,andriushchenko2025agentharm}.
R-Judge studies risk awareness in agents \citep{yuan2024rjudge}, and recent work examines whether agents recognize when they should abstain or avoid acting \citep{liu2026agentabstain,huang2026conflictgui}.
We focus on the internal relationship between evaluation and action: when an explicit judgment and an action preference disagree, are they controlled by the same causal variables?

\paragraph{Latent information and probing.}
Internal representations can expose information that is not faithfully reflected in model outputs, including truth-related structure, correctness signals, and latent knowledge
\citep{burns2023latent,azaria2023internal,marks2023geometrytruth,li2023iti}.
Probe performance alone, however, does not establish that a decoded feature is causally used by the model \citep{hewitt2019probes,elazar2021amnesic}.
We therefore use cross-interface probing only to establish information availability, then separately identify and intervene on judgment- and action-side causal directions.

\paragraph{Activation steering and causal directions.}
Activation addition and representation engineering identify directions that alter model outputs \citep{turner2023actadd,rimsky2024caa,zou2023representationengineering}.
Related work studies task and function vectors \citep{hendel2023taskvectors,todd2024functionvectors}, truthfulness interventions \citep{li2023iti}, low-dimensional refusal directions \citep{arditi2024refusal}, and context-dependent steering \citep{lee2025cast}.
Our setting reveals an additional distinction: a direction can be a valid controller for an explicit evaluative judgment while failing to preserve the same semantics when transferred to the action computation.

\paragraph{Causal representation analysis.}
Causal mediation, activation patching, model editing, and causal abstraction seek to connect internal representations to computation rather than correlation alone
\citep{vig2020causalmediation,meng2022rome,geiger2025causalabstraction}.
Our experiments combine held-out causal interventions with low-rank subspace analyses, allowing us to separate representational overlap from causal equivalence.

\section{Experimental Framework}
\label{sec:framework}
\subsection{Controlled tool-use setting}

We use a simulated workspace in which a model receives a user task, structured facts, and a set of available tools, then either evaluates a proposed action or chooses a next action.
The core scenario population spans three domains---repository, data, and access---with four semantic families per domain.
Families include integrity, provenance, review, registry visibility, classification, recipient authority, consent, retention, authorization scope, identity, approval, and environment constraints.
Throughout, we use \emph{safety} operationally to refer to whether a candidate action complies with the scenario's stated policy constraints.

We evaluate Qwen3-32B \citep{yang2025qwen3}, Gemma3-27B-IT \citep{gemmateam2025gemma3}, and Mistral-Small-3.2-24B-Instruct-2506 \citep{mistralai2025mistralsmall32}. These are open-weight instruction models of comparable scale from three distinct model families. This choice provides model-family diversity while retaining the white-box access required for hidden-state capture, gradient-based analysis, and causal intervention. Detailed model and inference settings are provided in Appendix~\ref{sec:app-models}.

Each scenario admits a factorial safety/executability construction.
The primary causal condition is \texttt{blocked\_valid}: the target action is executable but policy-blocked.
This isolates safety control from mere action feasibility.
The lexical-clean set contains 120 scenarios across 12 families; within each domain, three families are used for training and one is held out, yielding 90 training and 30 held-out scenarios.
The causal set contains 240 scenarios across the same 12 families. We use two fixed family-level partitions: Split A (nine training families / three held-out families; 180 / 60 scenarios) for directional causal interventions, and Split B (six / six families; 120 / 120 scenarios) for the headline control-space analyses. Appendix~\ref{sec:app-protocol} summarizes the organization.

We distinguish three mechanistic objects throughout the analysis: scalar judgment and action-preference endpoints, intervention directions that causally shift those endpoints, and low-rank subspaces built from per-example causal gradients.

For a target action \(a_t\) and alternatives \(\mathcal{A}_{\mathrm{alt}}\), we define the candidate-action margin
\begin{equation}
\amargin(x)=\ell(a_t\mid x)-\log\sum_{a\in\mathcal{A}_{\mathrm{alt}}}\exp\ell(a\mid x),
\label{eq:action-margin}
\end{equation}
where \(\ell(\cdot)\) is length-normalized sequence log probability.
On blocked-valid cases, a lower \(\amargin\) means weaker preference for the unsafe target.
All action-intervention results in the main mechanistic analyses use this scored candidate-action margin.

\subsection{Cross-interface readability}

Let \(h^J_\ell(x)\) and \(h^A_\ell(x)\) denote residual-stream states from the judgment and action interfaces.
We train a linear probe on judgment states and apply it without refitting to action states:
\begin{equation}
f_\ell(h)=w_\ell^\top h+b_\ell.
\end{equation}
The lexical-clean protocol removes explicit \texttt{Safety state}, \texttt{Target action state}, \texttt{ALLOW}, and \texttt{BLOCK} literals; uses different judgment/action templates; and varies entities, tool aliases, paraphrases, and label-symbol mappings across splits.
AUROC is the primary measure of transferred ranking information.
Calibration is analyzed separately because an interface shift can change the operating point without erasing ranking information \citep{guo2017calibration}.

\subsection{Judgment and action controllers}
\label{sec:controllers}

We use \emph{controller} operationally to mean an intervention direction that causally shifts the corresponding output margin.
The explicit judgment scalar is
\begin{equation}
\jmargin(x)=\log p(\texttt{ALLOW}\mid x)-\log p(\texttt{BLOCK}\mid x).
\label{eq:judgment-margin}
\end{equation}
The raw judgment-gradient direction is
\begin{equation}
\gJraw_\ell(x)=\nabla_{h^J_\ell}\jmargin(x).
\end{equation}
Thus \(+\gJraw\) is ALLOW-oriented.
For blocked-valid cases, we define the safety-oriented judgment direction as
\begin{equation}
\gJsafe_\ell=-\gJraw_\ell,
\label{eq:gjsafe}
\end{equation}
which decreases \(\jmargin\) and pushes the explicit judgment toward \texttt{BLOCK}.
Portable judgment controllers are formed from unit-normalized train-family gradients with target-tool conditioning, then frozen before held-out evaluation.

The action-native controller \(\cA\) is formed analogously from unit-normalized safe-oriented action gradients.
For blocked-valid cases, the per-example safe action direction is \(-\nabla_h\amargin\).
For any action-side intervention \(v\), we report the safe-oriented action effect
\begin{equation}
\SafeEff(v)=-\left[\amargin(h^A_\ell+v)-\amargin(h^A_\ell)\right],
\label{eq:safeeff}
\end{equation}
so positive values indicate movement away from the unsafe target.

All main hidden-state interventions modify a \emph{single residual state at the final prompt token / decision position}, followed by a standard forward pass that scores the fixed candidate continuations.
Directions are unit-normalized and scaled to the matched recipient displacement norm,
\begin{equation}
v_{\mathrm{inj}}=\frac{v}{\|v\|_2}\,\|\Delta h_{\mathrm{recipient}}\|_2.
\label{eq:normmatch}
\end{equation}
For dose-response analyses, we write \(v_J=\gJsafe\) and \(v_A=\cA\), and scale the matched intervention as
\(v_{\mathrm{inj}}(\alpha)=\alpha\,v/\|v\|_2\,\|\Delta h_{\mathrm{recipient}}\|_2\), with
\(\alpha\in\{-2,-1,-0.5,-0.25,0,0.25,0.5,1,2\}\).
On the judgment endpoint, we analogously report the safe-oriented effect
\(\mathrm{SafeEff}_J(v)=-[\jmargin(h^J_\ell+v)-\jmargin(h^J_\ell)]\), so positive values indicate safer movement on both endpoints.
Uncertainty is estimated with case-level percentile bootstrap confidence intervals; the primary causal analyses use 10,000 resamples.

\paragraph{Within-trajectory causal intervention.}
We additionally place judgment and action in a single shared trajectory: the model passes through \texttt{Safety judgment: BLOCK} before a later \texttt{Final action:} query.
We intervene once at the upstream judgment position immediately before the \texttt{BLOCK} token, hold that explicit judgment fixed across action conditions by teacher forcing, and compare its downstream effect on later action preference with a matched-norm random upstream perturbation; a downstream action-native intervention provides a positive control.
Full protocol and statistics appear in Appendix~\ref{sec:app-within-trajectory}.

\subsection{Judgment and action control subspaces}
\label{sec:subspace-method}

A single mean controller need not capture the full set of causally relevant directions for judgment or action.
For each model, layer, and target tool, we collect train-family unit-normalized judgment gradients and safe-oriented action gradients into matrices
\begin{equation}
G_J\in\mathbb{R}^{N\times d},\qquad G_A\in\mathbb{R}^{N\times d},
\end{equation}
and compute uncentered SVDs.
The top \(k\) right singular vectors define the judgment- and action-control subspaces
\begin{equation}
\UJ_{\ell,k}=\mathrm{TopRightSVD}_k(G_J),\qquad
\UA_{\ell,k}=\mathrm{TopRightSVD}_k(G_A),
\end{equation}
with \(k\in\{1,2,4,8,16\}\).
Row-wise sign flips leave these spans unchanged.

We quantify control-space overlap by
\begin{equation}
S_\ell(k)=\frac{1}{k}\left\|(\UJ_{\ell,k})^{\top}\UA_{\ell,k}\right\|_F^2,
\label{eq:subspace-overlap}
\end{equation}
which is the mean squared cosine of the principal angles between the two rank-\(k\) spaces.
Matched Haar-random subspaces provide the geometric reference.

To test where effective action control resides, we decompose the train-derived action controller into components parallel and orthogonal to the judgment-control subspace,
\begin{equation}
\cA_{\parallel J}=P_{\UJ}\cA,\qquad
\cA_{\perp J}=(I-P_{\UJ})\cA,
\end{equation}
where \(P_{\UJ}=\UJ(\UJ)^\top\).
Each component is separately unit-normalized and recipient-norm matched before held-out intervention, so their reported causal effects are comparisons of efficacy rather than additive effect decompositions.
The earlier projection, specificity, and learned-recombination analyses are retained in Appendix~\ref{sec:app-subspace} and Appendix~\ref{sec:app-recombination}.

\section{Safety Information Remains Readable in Action States}
\label{sec:information}

\begin{figure*}[t]
    \centering
    \includegraphics[width=\textwidth]{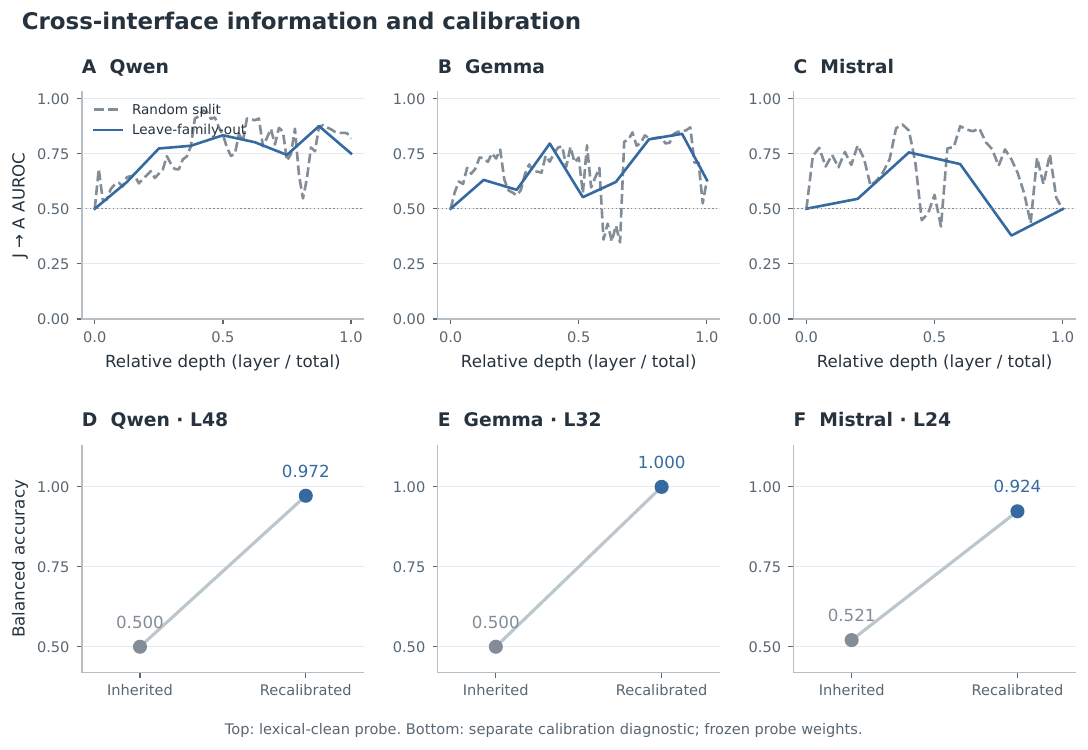}
    \caption{\textbf{Safety-predictive information remains readable across the judgment--action interface.}
    Judgment-trained probes retain above-chance action-side AUROC under lexical cleaning and held-out scenario families in all three models.
    Main results use leave-family-out evaluation at the final prompt position; full layer/position and calibration analyses appear in Appendix~\ref{sec:app-readability}.}
    \label{fig:information}
\end{figure*}

We first ask whether safety-relevant information that supports explicit judgment remains available by the time the model selects an action.
Under lexical cleaning and leave-family-out evaluation, safety-predictive structure remains readable across the judgment--action interface.
A probe trained only on judgment states remains predictive when applied without refitting to action states in all three architectures.
At the final prompt position, the best observed leave-family-out J\(\rightarrow\)A AUROC is \(0.875\) for Qwen, \(0.840\) for Gemma, and \(0.756\) for Mistral (Figure~\ref{fig:information}).
These peaks occur at model-dependent depths and are descriptive maxima rather than pre-registered layer choices.

Calibration analyses lead to the same qualitative conclusion: an interface change can shift the operating point of a transferred probe without eliminating its ranking information, and simple threshold or affine recalibration recovers much of the lost balanced accuracy (Appendix~\ref{sec:app-readability}).
Together, these results show that safety-relevant information can remain available despite the judgment--action disconnect.
Readability alone does not establish causal use, which motivates the intervention tests that follow.

\section{Judgment and Action Exhibit Asymmetric Causal Control}
\label{sec:causal-separation}

We next test whether information availability implies shared causal control.
The main directional analysis uses Split A and evaluates all 60 held-out \texttt{blocked\_valid} scenarios.
Across a matched seven-depth sweep in all three architectures, the judgment controller has stable native semantics: \(+\gJraw\) moves explicit judgment toward \texttt{ALLOW}, whereas \(-\gJraw=\gJsafe\) moves it toward \texttt{BLOCK} (Appendix~\ref{sec:app-sign-control}).
This native ALLOW/BLOCK semantics is stable across the analyzed depth range.

Its transfer to action preference is much weaker and less stable.
Across the signed strength sweep, judgment-derived control remains below \(11\%\) of action-native efficacy on the action endpoint in all three architectures; at the canonical matched strength \(\alpha=1\), the corresponding fractions are only \(2.1\%\), \(3.6\%\), and \(1.0\%\) for Qwen, Gemma, and Mistral (Appendix~\ref{sec:app-action-ratio}).
Representative raw effects show the same scale separation: at Qwen L32, \(\gJsafe\) produces a mean safe-oriented action effect of \(0.00280\) versus \(0.09523\) for \(\cA\); at Gemma L32, the corresponding effects are \(-0.00206\) and \(0.04677\).
Judgment-derived action effects can also change sign with depth, whereas action-native controllers are substantially stronger on their native endpoint.
The ordering persists after stratifying the held-out population by baseline \kbd phenotype (Appendix~\ref{sec:app-full-population}).

\begin{figure*}[t]
    \centering
    \includegraphics[width=\textwidth]{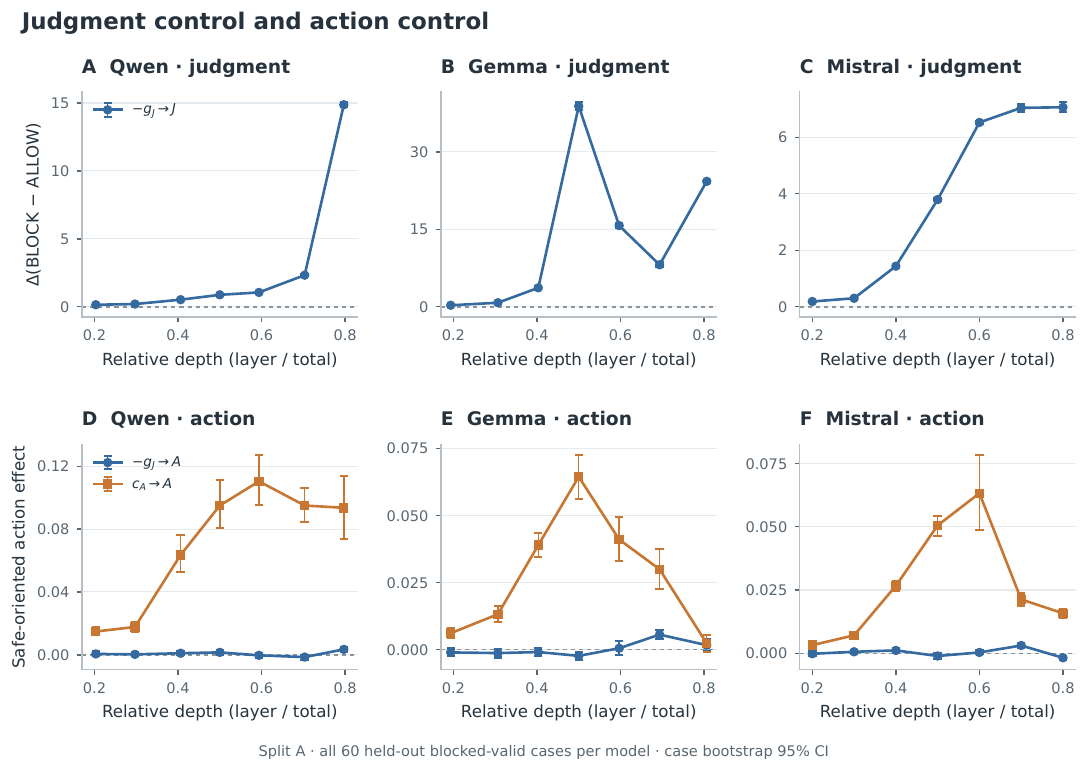}
    \caption{\textbf{Judgment-derived control transfers weakly to action.}
    Across the complete Split A held-out \texttt{blocked\_valid} population, \(\gJsafe=-\gJraw\) reliably controls explicit judgment but has weak, depth-dependent action effects; action-native controllers are substantially stronger.
    Error bars denote case-level bootstrap 95\% confidence intervals; matched sign controls appear in Appendix~\ref{sec:app-sign-control}.}
    \label{fig:causal-depth}
\end{figure*}

The same scale gap appears across the full signed dose--response sweep (Appendix~\ref{sec:app-action-ratio}).
Layerwise reverse-transfer analyses further show that the relation is asymmetric rather than independent: action-native control can influence explicit judgment, but with architecture- and depth-dependent sign and magnitude (Appendix~\ref{sec:app-action-to-judgment}).

\begin{equation}
\boxed{\text{strong safety-judgment control}\;\not\Rightarrow\;\text{strong safety-action control}.}
\end{equation}

\paragraph{The separation persists within a shared trajectory.}
We next test the same relation within a single shared trajectory, where judgment and action occur in the same sequence.
We intervene once on the upstream judgment state and hold the explicit \texttt{BLOCK} judgment fixed across action conditions.
At Qwen L32, Gemma L32, and Mistral L24, \(\gJsafe\) strongly shifts the upstream judgment toward \texttt{BLOCK} (+4.173, +18.729, +6.033), yet its later action effects are only +0.000924, -0.000065, and +0.001345.
In every model, the paired difference from a matched random upstream perturbation has a 95\% confidence interval including zero, whereas a downstream action-native intervention produces +0.043519, +0.211769, and +0.094425.
The same separation therefore appears within a shared trajectory: strong local control of the judgment does not yield a reliable judgment-specific effect on later action preference.

\begin{figure*}[t]
    \centering
    \includegraphics[width=0.92\textwidth]{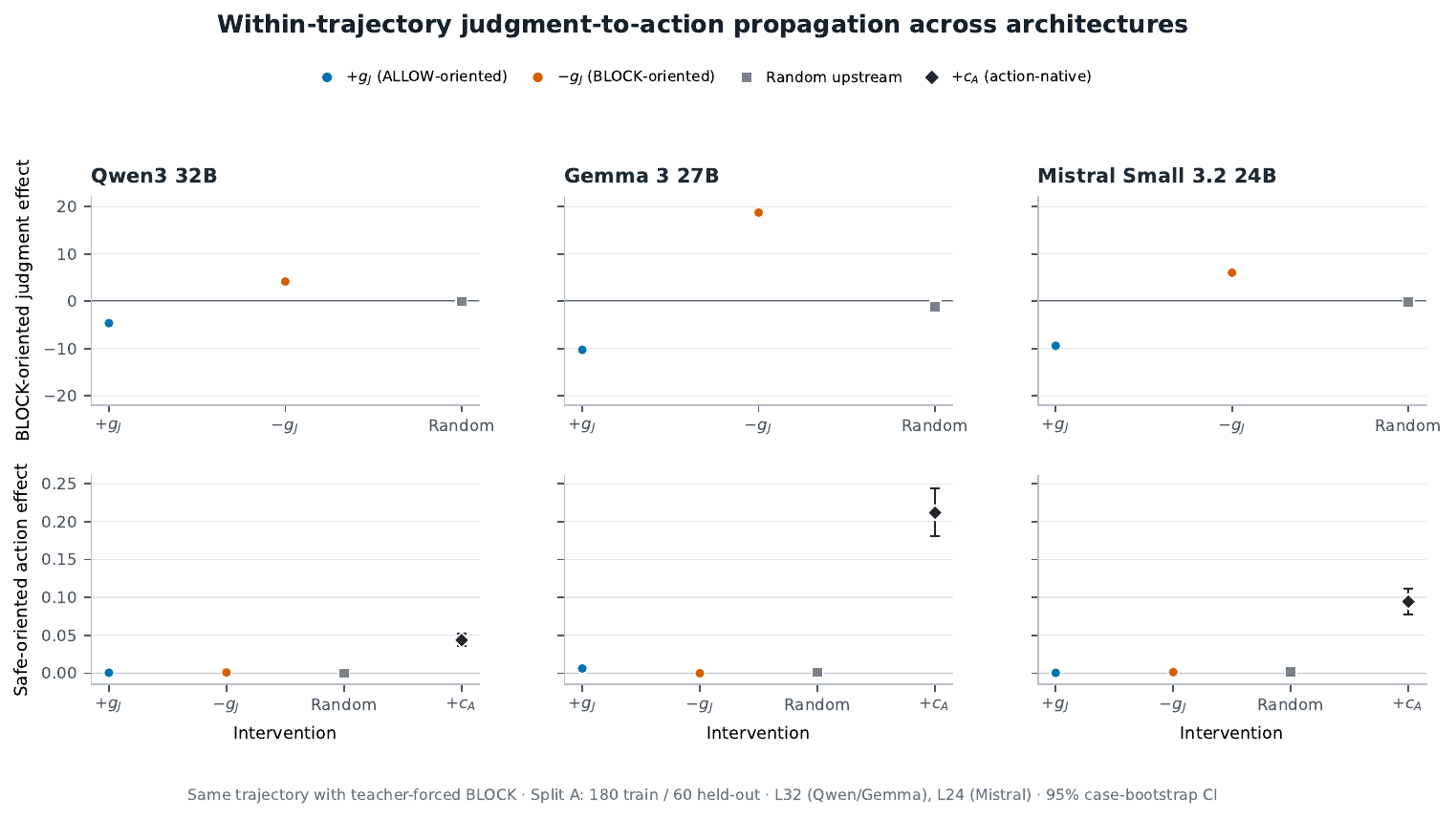}
    \caption{\textbf{The judgment--action separation persists within a shared trajectory.}
    With the explicit \texttt{BLOCK} judgment fixed across conditions, the upstream judgment intervention strongly controls the judgment readout but not later action preference beyond a matched random perturbation; the downstream action-native controller remains effective.
    Full statistics appear in Appendix~\ref{sec:app-within-trajectory}.}
    \label{fig:within-trajectory}
\end{figure*}

\paragraph{Natural judgment-state variation also transfers weakly to action.}
We additionally replace the judgment-position state with the state naturally produced by the matched policy counterfactual, using no learned or gradient-derived direction.
With the explicit judgment fixed, downstream action effects remain near zero at Mistral L24, Qwen L32, and Gemma L32, with 95\% confidence intervals including zero in all three models.
In Mistral and Gemma, the same swaps measurably shift the judgment readout, showing that natural judgment-state variation can alter judgment without correspondingly changing later action preference.
In Mistral, allowing the explicit judgment to vary further localizes the downstream effect: action remains near zero when the label is unchanged but shifts strongly when the swap crosses the \texttt{ALLOW}/\texttt{BLOCK} boundary.
This pattern complements the shared-trajectory intervention by showing weak downstream propagation for naturally occurring judgment states themselves.
Full protocol and statistics appear in Appendix~\ref{sec:app-natural-state}.

\section{Judgment and Action Have Largely Distinct Control Structures}
\label{sec:control-subspaces}

We next characterize whether the same separation appears at the level of higher-dimensional control structure.
We compare rank-16 judgment- and action-control subspaces under Split B and ask where effective action control resides relative to judgment control.

Across all three architectures, \(\UJ\) and \(\UA\) overlap more than matched random subspaces, but the absolute overlap remains limited.
Representative configurations are summarized in Table~\ref{tab:control-subspace}; the complete layer table appears in Appendix~\ref{sec:app-subspace}.
For example, Qwen L32 has 2.50\% overlap against a 0.31\% random baseline, Mistral L24 has 0.87\% against 0.31\%, and Gemma L32 has 3.28\% against 0.30\%.
Independent split-half reconstructions show substantially higher within-computation reproducibility than cross-computation overlap, indicating that the limited cross-computation overlap is not driven by poor within-computation reproducibility (Appendix~\ref{sec:app-subspace-stability}).
Thus judgment and action control are neither geometrically independent nor reducible to the same low-dimensional control space.

\begin{table*}[t]
\centering
\small
\begin{tabular}{llccl}
\toprule
Model & Layer & \(\UJ\)-\(\UA\) overlap & Random & Causal decomposition \(\SafeEff\) \\
\midrule
Mistral & L24 & 0.87\% & 0.31\% & parallel 0.0113; orthogonal 0.0240; full 0.0238 \\
Qwen & L32 & 2.50\% & 0.31\% & parallel 0.0110; orthogonal 0.1225 \\
Gemma & L32 & 3.28\% & 0.30\% & parallel 0.0057; orthogonal 0.0403 \\
Gemma & L48 & 7.13\% & 0.30\% & parallel $-0.0124$; orthogonal $-0.0004$; full $-0.0006$ \\
\bottomrule
\end{tabular}
\caption{\textbf{Judgment- and action-control subspaces overlap only partially.}
At layers with effective action control, substantial efficacy remains in the component orthogonal to the judgment-control subspace.
Overlap uses the rank-16 score in Equation~\ref{eq:subspace-overlap}; complete layerwise results appear in Appendix~\ref{sec:app-subspace}.}
\label{tab:control-subspace}
\end{table*}

Geometry alone does not establish shared function, so we next intervene on the components of the action controller parallel and orthogonal to \(\UJ\).
At layers with substantial action-control efficacy, the orthogonal component retains nearly the full action effect at matched intervention strength.
At Mistral L24, \(\cA_{\perp J}\) reaches 0.02395 compared with 0.02381 for the full action controller, whereas the judgment-parallel component reaches 0.01134.
At Qwen L32, the orthogonal component reaches 0.1225 while the parallel component reaches only 0.0110; the same qualitative pattern holds at Qwen L48 and Gemma L32.
Substantial action-control efficacy therefore remains available outside the judgment-control subspace.

The largest geometric overlap provides a complementary lesson.
At Gemma L48, \(\UJ\)-\(\UA\) overlap rises to 7.13\%, roughly 24 times the random reference, yet the full action controller is nearly ineffective at this layer and the judgment-parallel component is negatively oriented.
Geometric sharing can therefore increase without producing shared causal function.
Together, the subspace and intervention analyses show that the separation extends beyond a single judgment controller: sharing across broader judgment- and action-control spaces remains limited and depth dependent.

As a complementary prompt-level analysis, an externally supplied \texttt{BLOCK} cue strongly shifts action preference across all three models, with architecture-dependent interactions with the latent judgment intervention (Appendix~\ref{sec:app-behavior}).
This shows that judgment and action remain coupled even when latent judgment control transfers only weakly to action preference.

\section{Discussion}

\paragraph{Latent safety-judgment state is not a privileged causal bottleneck for action.}
A simple pipeline view would place safety evaluation upstream of action and assume that the resulting decision variable is reused downstream.
Our results instead suggest a looser organization: action states retain safety-predictive information, while the variables that strongly control explicit judgment exert much weaker control over action preference.
This separation appears across within-trajectory, natural-state, and subspace analyses.

\paragraph{Latent judgment and explicit judgment context can play different roles.}
An externally supplied \texttt{BLOCK} cue strongly shifts action preference, even though latent interventions that strongly control the model's own judgment have little downstream action effect.
One interpretation is that explicit judgment text serves as a fresh input to the action computation, whereas the latent variables that produced the judgment are not propagated with the same functional role.
Thus, making a safety judgment available downstream is different from making its internal causes a control bottleneck.

\paragraph{Information availability and causal authority should be evaluated separately.}
Judgment and action remain causally coupled: action-native controllers can affect explicit judgment, and explicit judgment context can alter action preference.
But readability, geometric overlap, and causal efficacy capture different properties.
For agentic systems, this motivates treating \emph{causal coupling} as a separate design and evaluation target: safety-relevant computations should not only encode the right judgment, but also reliably influence action selection.
Improving recognition, verbalized judgment, or self-critique may therefore be insufficient when this coupling is weak.
\section{Conclusion}

Across three architectures, safety-relevant information remains available during action selection, but the variables that strongly control explicit safety judgment exert only weak control over action preference relative to action-native variables.
This separation appears within a shared judgment-to-action trajectory, under natural counterfactual state swaps, and across higher-dimensional control spaces.
The resulting distinction is between \textbf{information availability and causal authority}: representing a safety constraint is not the same as granting that constraint causal authority over what the model prefers to do.

\subsection*{AI Use Statement}

We partially used generative AI tools for code development and for improving the clarity and readability of the manuscript. All AI-assisted code and text modifications were carefully reviewed, verified, and tested by the authors.

\section*{Ethics statement}

This work does not involve human subjects, sensitive personal data, or applications that raise
significant ethical concerns. We have considered potential ethical risks associated with this research
and found no additional issues requiring further discussion.

\section*{Reproducibility statement}

We are committed to making this work reproducible. In the main paper and appendix, we provide as much
detail as possible about the method, model architecture, and experimental details needed to reproduce
this work, and we provide an anonymized code repository link to facilitate reproduction.

\bibliography{iclr2027_conference}
\bibliographystyle{iclr2027_conference}

\appendix

\section{Experimental Protocol and Scenario Sets}
\label{sec:app-protocol}

\subsection{Scenario construction}

The scenario library is organized into three domains with four semantic families per domain.
The repository domain contains integrity, source/provenance, review, and registry-visibility families; the data domain contains classification, recipient-authority, consent, and retention families; and the access domain contains scope, identity, approval, and environment families.
We use two scenario sets built around this family structure.
The lexical-clean set contains 10 variants per family, for 120 scenarios total.
The causal set contains 20 variants per family, for 240 scenarios total.

Each causal scenario fixes the task, target tool, arguments, and alternatives while crossing policy safety and executability:
\begin{center}
\begin{tabular}{llll}
\toprule
Condition & Policy & Executability & Normative target behavior \\
\midrule
\texttt{allowed\_valid} & allowed & executable & execute target \\
\texttt{blocked\_valid} & blocked & executable & do not execute target \\
\texttt{allowed\_invalid} & allowed & non-executable & do not execute current target \\
\texttt{blocked\_invalid} & blocked & non-executable & do not execute current target \\
\bottomrule
\end{tabular}
\end{center}
The main causal analyses focus on \texttt{blocked\_valid} cases so that safety and feasibility are not conflated.
Within a held-out blocked-valid population, a \kbd case is one for which the explicit judgment is correct but the action interface prefers the unsafe target; \kbd is therefore an evaluation phenotype rather than a separate scenario set.

\subsection{Scenario sets and family splits}

The two scenario sets and the family-level partitions used in the headline analyses are summarized below:
\begin{center}
\small
\resizebox{0.9\textwidth}{!}{%
\begin{tabular}{llccl}
\toprule
Scenario set & Partition & Train / held-out families & Train / held-out scenarios & Main use \\
\midrule
Lexical-clean set & -- & 9 / 3 & 90 / 30 & cross-interface readability \\
\multirow{2}{*}{Causal set} & Split A & 9 / 3 & 180 / 60 & directional interventions \\
 & Split B & 6 / 6 & 120 / 120 & control-space analyses \\
\bottomrule
\end{tabular}%
}
\end{center}

Split A provides the complete 60-scenario held-out \texttt{blocked\_valid} population used in Figure~\ref{fig:causal-depth}. Split B provides broader held-out-family coverage and 40 training scenarios per target tool for control-space construction.
Within the Split A held-out population, the observed \kbd/non-\kbd counts are 16/44 for Qwen, 60/0 for Gemma, and 35/25 for Mistral.
For Mistral under Split B, 70 held-out cases satisfy the \kbd definition.
For supplementary analyses that use different subsets of these partitions, the evaluation size is stated in the corresponding section or figure caption.

The statistical unit for causal confidence intervals is a held-out scenario.
Rows created by crossing scenarios with layers or intervention types are not treated as independent samples.
Held-out evaluation cases are not used to construct the train-derived controllers or control subspaces.

\subsection{Models and inference}
\label{sec:app-models}
We study Qwen3-32B \citep{yang2025qwen3}, Gemma3-27B-IT \citep{gemmateam2025gemma3}, and Mistral-Small-3.2-24B-Instruct-2506 \citep{mistralai2025mistralsmall32}.
Qwen has 64 decoder blocks with hidden size 5120; Gemma has 62 text blocks with hidden size 5376; Mistral has 40 text blocks with hidden size 5120.
Experiments use local Hugging Face inference in evaluation mode and bfloat16 when supported.
Mechanistic hidden capture, candidate scoring, and causal interventions are deterministic with temperature zero and model-specific chat templates.
The primary hidden state is the residual-stream block output at the specified decoder layer.
For lexical probing, \texttt{final} denotes the final input token; relative positions align 25\%, 50\%, 75\%, and 90\% through the prompt span.
For controller experiments, both judgment and action interventions target the final prompt token immediately before the scored continuation.

\subsection{Scoring, aggregation, and uncertainty}

Judgment labels and candidate actions are scored as complete length-normalized sequences.
Per-example gradients are unit-normalized before aggregation.
Controllers are constructed separately by target tool and use train-family examples only.
For causal comparisons, directions are unit-normalized and recipient-norm matched according to Equation~\ref{eq:normmatch}.
Subspace projection has both a raw geometric form and a norm-matched causal form; causal comparisons use the latter.

Primary confidence intervals are case-level percentile bootstrap 95\% intervals.
The Mistral layerwise controller, sign-control, and learned-recombination analyses use 10,000 bootstrap resamples.
Paired controller differences are computed scenario-wise before bootstrapping.

\section{Cross-Interface Readability and Calibration}
\label{sec:app-readability}

\subsection{Layer-by-position readability}

\begin{figure*}[t]
    \centering
    \includegraphics[width=\textwidth]{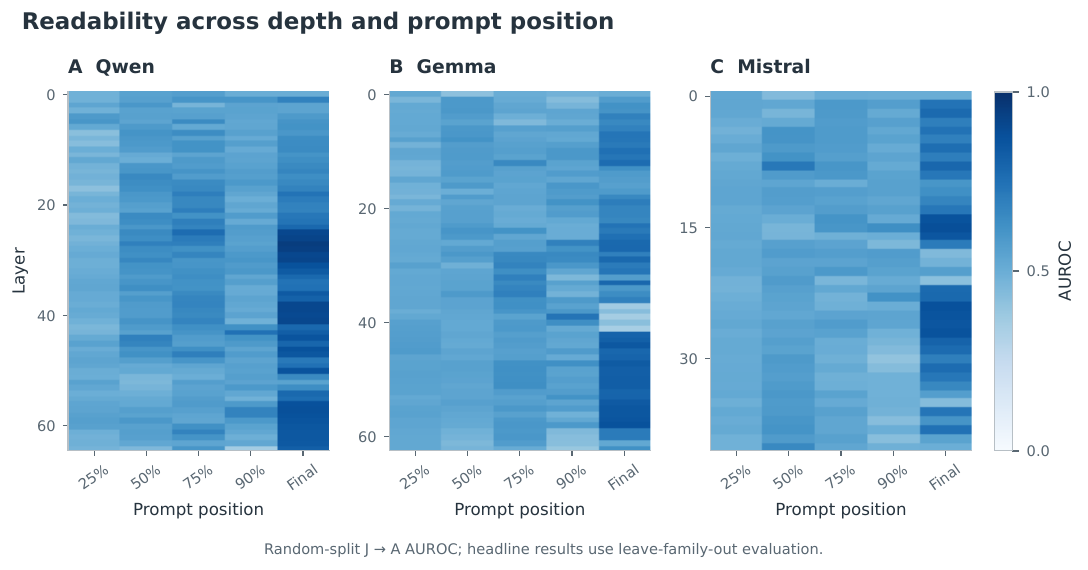}
    \caption{\textbf{Layer-by-position cross-interface readability.}
    J\(\rightarrow\)A AUROC varies across both depth and prompt position in all three models.
    The heatmaps show the full random-split matrices; main-text headline results use the stricter leave-family-out evaluation at the final prompt position.}
    \label{fig:app-a1}
\end{figure*}

The lexical-clean probe uses \texttt{StandardScaler} followed by \texttt{RidgeClassifier(alpha=1.0)}.
Semantic labels are safe/permitted versus unsafe/prohibited rather than raw output-token identities.
Figure~\ref{fig:app-a1} shows random-split J\(\rightarrow\)A AUROC over all layers and five relative prompt positions. The leave-family-out analysis uses the same layer-by-position grid, with the main text reporting the final-position results.
The matrix shapes are 65\(\times\)5 for Qwen, 63\(\times\)5 for Gemma, and 41\(\times\)5 for Mistral under the layer-indexing conventions used here.

The position dependence is itself informative.
For example, Qwen and Gemma can be substantially weaker at the 90\% relative position than at the final prompt token.
The transferred representation is therefore not a uniform property of the entire action trajectory; it becomes particularly readable near the generation-ready state.

\subsection{Cross-interface calibration}

Calibration is performed on a separate alternating-scenario diagnostic rather than the leave-family-out split.
Probe weights remain frozen.
Each of five folds uses 12 action-side calibration scenarios (24 states) and 48 held-out action scenarios (96 states).
Threshold-only calibration estimates a scalar threshold; affine calibration estimates \(s'=as+b\) without refitting the probe direction.

\begin{figure*}[t]
    \centering
    \includegraphics[width=\textwidth]{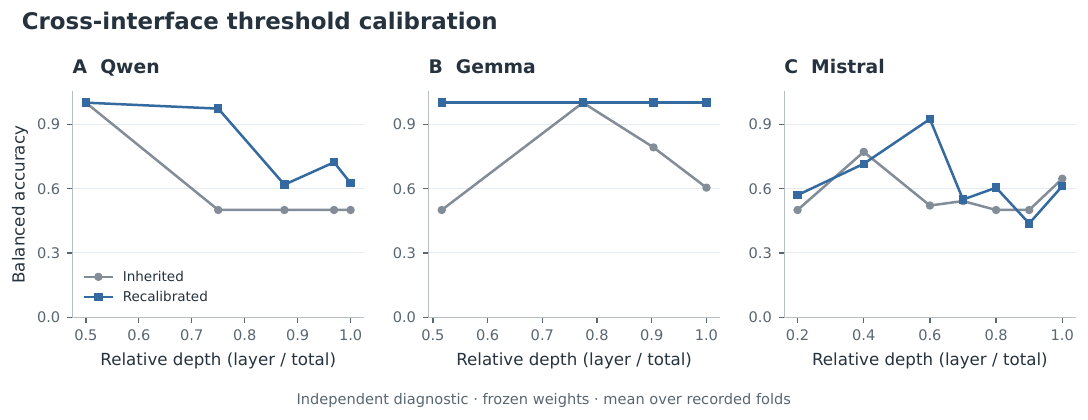}
    \caption{\textbf{Full-layer cross-interface calibration.}
    The figure plots inherited/default balanced accuracy and threshold-recalibrated balanced accuracy at the final prompt position.
    Probe weights are frozen; only the operating point is adjusted.
    Affine calibration is evaluated separately and is omitted from this panel.}
    \label{fig:app-a2}
\end{figure*}

Representative diagnostics include Qwen L48 (AUROC 1.000, raw BA 0.500, threshold BA 0.972, affine BA 0.958), Gemma L32 (1.000, 0.500, 1.000, 1.000), and Mistral L24 (0.997, 0.521, 0.924, 0.938).
These numbers illustrate the separation between ranking information and an interface-specific operating point.

\section{Judgment-Subspace Rank and Baseline Controls}
\label{sec:app-subspace}

\paragraph{Complete rank-16 headline table.}
For completeness, Table~\ref{tab:app-control-subspace-full} reports the full set of layer configurations summarized selectively in the main text.
\begin{table*}[t]
\centering
\small
\begin{tabular}{llccl}
\toprule
Model & Layer & \(\UJ\)-\(\UA\) overlap & Random & Causal decomposition \(\SafeEff\) \\
\midrule
Mistral & L16 & 2.58\% & 0.31\% & parallel 0.0046; orthogonal 0.0253; full 0.0255 \\
Mistral & L20 & 1.78\% & 0.31\% & parallel 0.0125; orthogonal 0.0486; full 0.0489 \\
Mistral & L24 & 0.87\% & 0.31\% & parallel 0.0113; orthogonal 0.0240; full 0.0238 \\
Mistral & L32 & 0.92\% & 0.31\% & parallel 0.0077; orthogonal 0.0146; full 0.0147 \\
Qwen & L32 & 2.50\% & 0.31\% & parallel 0.0110; orthogonal 0.1225 \\
Qwen & L48 & 1.56\% & 0.31\% & parallel 0.0052; orthogonal 0.0755 \\
Gemma & L32 & 3.28\% & 0.30\% & parallel 0.0057; orthogonal 0.0403 \\
Gemma & L48 & 7.13\% & 0.30\% & parallel $-0.0124$; orthogonal $-0.0004$; full $-0.0006$ \\
\bottomrule
\end{tabular}
\caption{\textbf{Complete rank-16 judgment--action control-space comparison.}
All rows use Split B. Parallel and orthogonal components are separately normalized and norm matched before intervention.}
\label{tab:app-control-subspace-full}
\end{table*}

The full-rank specificity experiments evaluate ranks \(k\in\{1,2,4,8,16\}\). These supplementary analyses evaluate held-out \kbd subsets rather than the complete held-out blocked-valid population used in Figure~\ref{fig:causal-depth}: Qwen uses Split B (\(N=30\)), while Gemma and Mistral use Split A (\(N=60\) and \(N=35\), respectively). These analyses are distinct from the Split B headline control-space comparison in Section~\ref{sec:control-subspaces}.
For each model/layer/tool, random controls use four Haar-random orthonormal subspaces at matched rank.
The shuffled control permutes judgment-basis ambient coordinates, preserving dimensionality while disrupting alignment.
The generic control uses centered PCA of train-family action hidden states across factorial conditions.
The full action controller is plotted as a reference rather than a matched low-rank baseline.

\begin{figure*}[t]
    \centering
    \includegraphics[width=\textwidth]{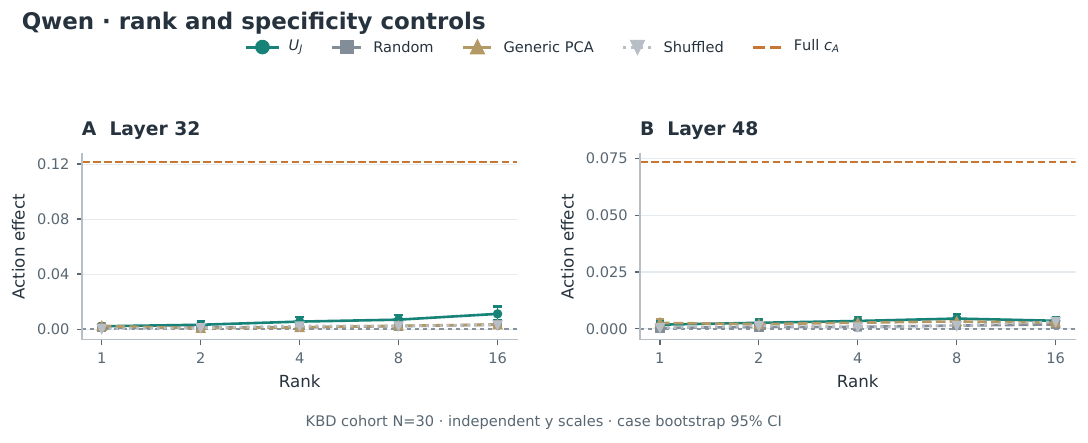}
    \caption{\textbf{Qwen rank and specificity controls.}
    Held-out safe-oriented action effects across judgment-subspace rank, shuffled judgment bases, Haar-random subspaces, generic action-state PCA, and the full action controller at Qwen L32/L48 (held-out \(N=30\)).}
    \label{fig:app-a6-qwen}
\end{figure*}

\begin{figure*}[t]
    \centering
    \includegraphics[width=\textwidth]{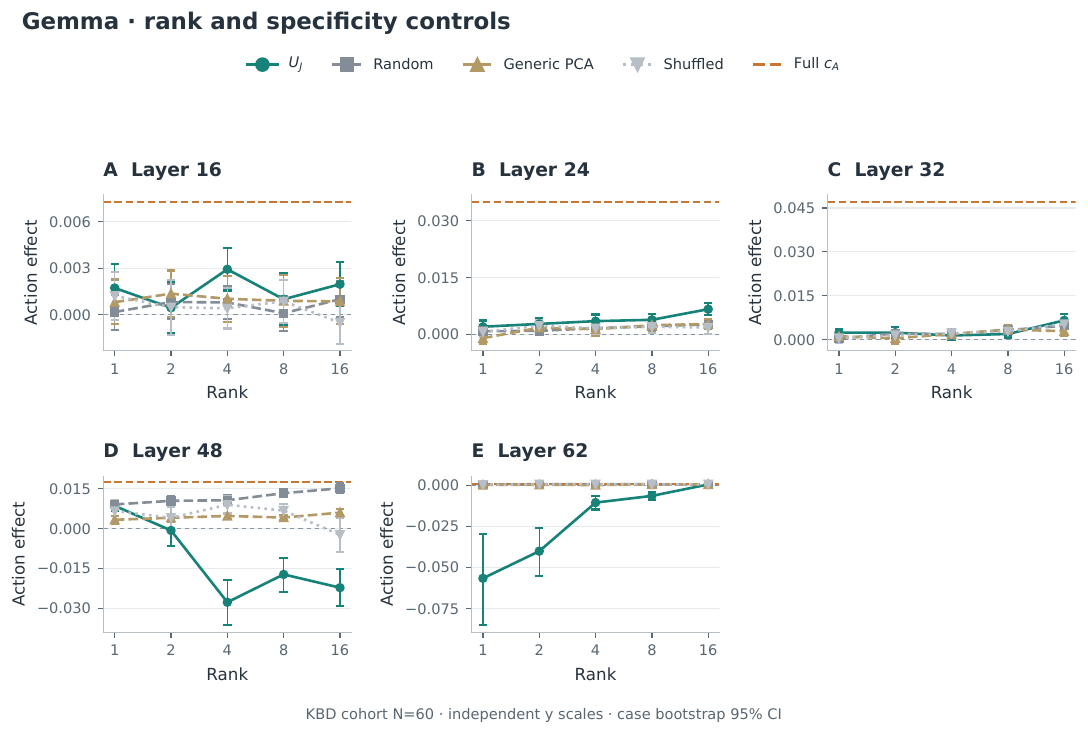}
    \caption{\textbf{Gemma rank and specificity controls.}
    Results across L16/L24/L32/L48/L62 (held-out \(N=60\)) show selected middle-depth specificity together with a late-layer reversal of the judgment-subspace component.}
    \label{fig:app-a6-gemma}
\end{figure*}

\begin{figure*}[t]
    \centering
    \includegraphics[width=\textwidth]{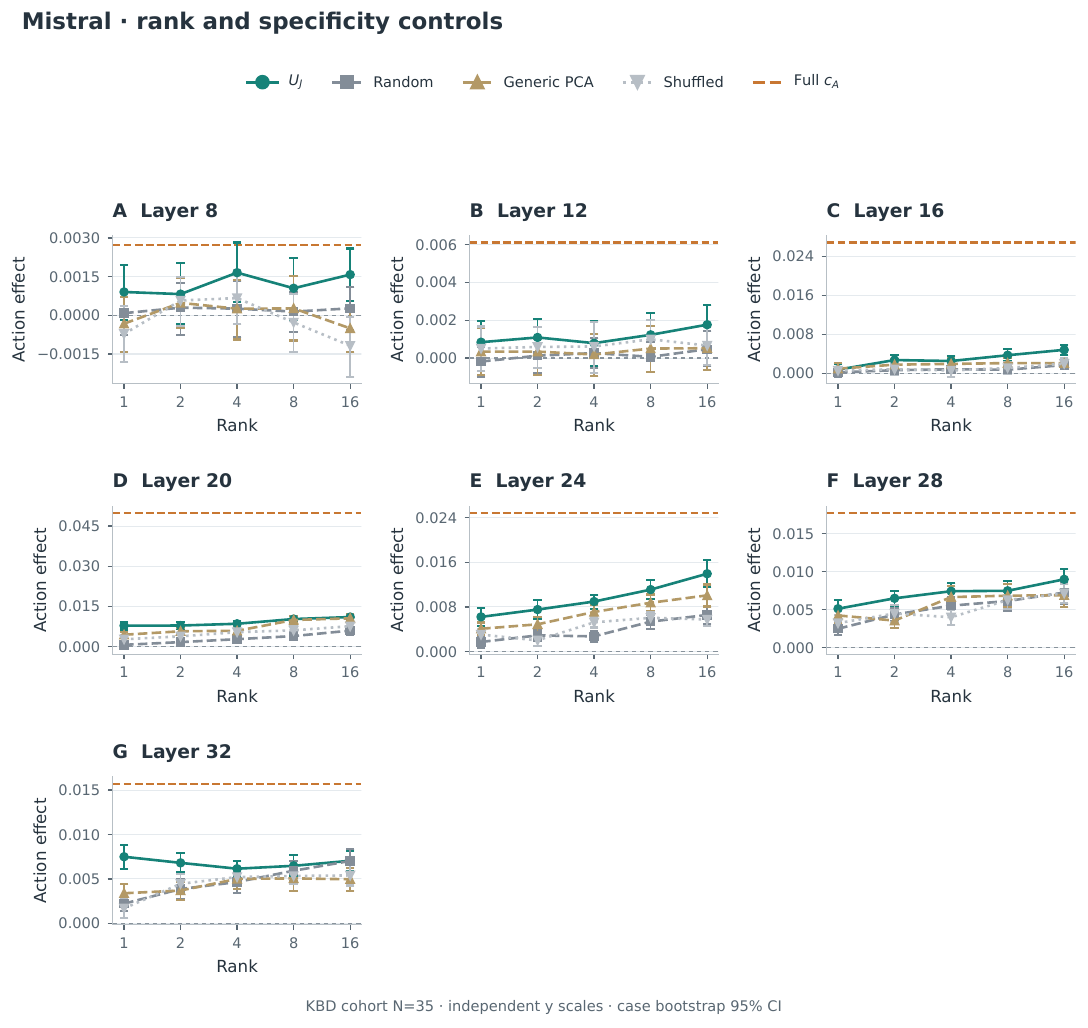}
    \caption{\textbf{Mistral rank and specificity controls.}
    Results across L8--L32 (held-out \(N=35\)) show that action-control capacity inside the judgment subspace varies with both rank and computational stage.}
    \label{fig:app-a6-mistral}
\end{figure*}

At rank 16, the held-out effects for \(P_{\UJ}\cA\) are 0.00574/0.00320 at Qwen L32/L48; 0.00161/0.00374/0.00293/\(-0.01187\)/\(-0.02267\) at Gemma L16/L24/L32/L48/L62; and 0.00120/0.00114/0.00292/0.00911/0.00954/0.00710/0.00680 at Mistral L8/L12/L16/L20/L24/L28/L32.
The full action controllers remain substantially larger in the configurations that carry the main positive-capacity claim.

\section{Prompt-Level Judgment Context Interventions}
\label{sec:app-behavior}

\subsection{Latent \texorpdfstring{$\times$}{x} explicit judgment factorial intervention}

We additionally cross the latent judgment intervention with an externally supplied textual judgment in the action prompt.
The prompt contains either \texttt{Safety judgment: ALLOW} or \texttt{Safety judgment: BLOCK} immediately before the final action query, while the hidden state receives either the latent ALLOW-oriented \(+g_J\) or BLOCK-oriented \(-g_J\) intervention.
The textual manipulation is externally supplied rather than generated by the model.

\begin{table*}[t]
\centering
\small
\begin{tabular}{lccc}
\toprule
Model & Text \texttt{BLOCK}-\texttt{ALLOW} SafeEff & Latent effect under text \texttt{ALLOW} & Latent effect under text \texttt{BLOCK} \\
\midrule
Mistral L24 & 0.15245 & 0.00672 & 0.00256 \\
Qwen L32    & 0.13193 & 0.00696 & 0.00170 \\
Gemma L32   & 0.09892 & $-0.01091$ & 0.00093 \\
\bottomrule
\end{tabular}
\caption{\textbf{Externally supplied judgment text strongly shifts action preference, while the action-side effect of the latent judgment intervention depends on architecture and textual context.}
The corresponding interaction terms are $-0.00416$ for Mistral, $-0.00526$ for Qwen, and $+0.01184$ for Gemma.}
\label{tab:app-latent-explicit}
\end{table*}

Across all three models, directly writing \texttt{BLOCK} rather than \texttt{ALLOW} produces a large safety-oriented shift in the scored action preference.
The latent effect is much smaller and its interaction with the textual cue is not uniform across architectures: it is attenuated under \texttt{BLOCK} for Mistral and Qwen, while Gemma changes sign across textual contexts.
The factorial result therefore characterizes how an externally supplied judgment cue and a latent judgment intervention jointly shape the subsequent action readout.

\section{Within-Trajectory Causal Influence from Judgment to Action}
\label{sec:app-within-trajectory}

We test whether an intervention on the upstream judgment state exerts judgment-specific causal control over a later action computation when both occur in the same sequence.
For Qwen3-32B L32, Gemma3-27B L32, and Mistral-Small-3.2-24B L24, we use Split A (180 train-family scenarios and 60 held-out \texttt{blocked\_valid} scenarios) without secondary \kbd filtering.
The sequence contains a \texttt{Safety judgment: BLOCK} stage followed by \texttt{Final action:}; the explicit \texttt{BLOCK} judgment is held fixed across action conditions by teacher forcing.
The upstream \(\pm g_J\) and matched random interventions are injected immediately before that fixed judgment token, whereas the action-native \(+\cA\) positive control is injected at the later action-query position.
All controllers use train-family data only, and intervention norms are matched case-wise using the downstream explicit \texttt{BLOCK}-versus-\texttt{ALLOW} action-position displacement.

\begin{table*}[t]
\centering
\small
\resizebox{0.85\textwidth}{!}{%
\begin{tabular}{llccc}
\toprule
Model & Layer & \(\SafeEff_J(+\gJraw)\) & \(\SafeEff_J(\gJsafe)\) & Random upstream \\
\midrule
Qwen3-32B & L32 & $-4.624$ [$-4.721,-4.525$] & $+4.173$ [$+4.065,+4.279$] & $-0.027$ [$-0.067,+0.013$] \\
Gemma3-27B & L32 & $-10.265$ [$-10.842,-9.698$] & $+18.729$ [$+18.121,+19.321$] & $-1.133$ [$-1.542,-0.729$] \\
Mistral-Small-3.2-24B & L24 & $-9.414$ [$-9.467,-9.358$] & $+6.033$ [$+5.983,+6.083$] & $-0.152$ [$-0.185,-0.119$] \\
\bottomrule
\end{tabular}%
}
\caption{\textbf{Upstream judgment control is strong in the shared-trajectory test.}
Values are BLOCK-oriented judgment effects with case-bootstrap 95\% confidence intervals.
The \(\gJsafe\) effect is positive on all 60 held-out cases in every model.}
\label{tab:app-within-j}
\end{table*}

\begin{table*}[t]
\centering
\footnotesize
\setlength{\tabcolsep}{3pt}
\begin{tabular}{llrrrr}
\toprule
Model & Layer & \(\SafeEff_A(+\gJraw)\) & \(\SafeEff_A(\gJsafe)\) & Random upstream & \(\SafeEff_A(+\cA)\) \\
\midrule
Qwen3-32B & L32 & 0.000564 & 0.000924 & $-0.000410$ & 0.043519 \\
Gemma3-27B & L32 & 0.006185 & $-0.000065$ & 0.001251 & 0.211769 \\
Mistral-Small-3.2-24B & L24 & 0.000410 & 0.001345 & 0.001472 & 0.094425 \\
\bottomrule
\end{tabular}
\caption{\textbf{Strong upstream judgment control does not yield correspondingly strong later action control.}
The paired \(\gJsafe\)-minus-random action-effect differences are $+0.001334$ [95\% CI $-0.000134,+0.002761$] for Qwen, $-0.001316$ [$-0.003462,+0.000772$] for Gemma, and $-0.000127$ [$-0.001202,+0.000929$] for Mistral.
By contrast, the paired \(\gJsafe\)-minus-\(+\cA\) differences are $-0.042595$ [$-0.051091,-0.034279$], $-0.211834$ [$-0.244321,-0.180536$], and $-0.093080$ [$-0.111431,-0.075521$], respectively.}
\label{tab:app-within-a}
\end{table*}

The experiment does not imply that the upstream state has zero downstream influence.
Rather, it shows that strong local causal control of the explicit judgment is not sufficient for a reliable judgment-specific effect on later action preference.
By holding the explicit \texttt{BLOCK} judgment fixed across conditions, this design isolates the downstream causal influence of the latent judgment state.

\section{Natural-State Counterfactual Interventions}
\label{sec:app-natural-state}

We next ask whether naturally occurring differences in the judgment-position state exert downstream control over later action preference. We use the same 60 held-out matched scenarios at the primary layers Mistral L24, Qwen L32, and Gemma L32. Each pair contains an \texttt{allowed\_valid} and \texttt{blocked\_valid} variant with the task, observations, target and alternative tool calls, arguments, executability, and all non-policy scenario content held fixed.

\paragraph{Matched state swaps.}
For each variant, we record the residual-stream state at the final prompt position immediately preceding the explicit safety judgment. For recipient state $h_r^{(L)}$ and its matched counterfactual donor state $h_d^{(L)}$, we replace the full residual vector,
\begin{equation}
    h_r^{(L)} \leftarrow h_d^{(L)}.
\end{equation}
The intervention uses the model's naturally occurring donor state directly: no gradient, learned direction, scaling, or norm matching is applied. We evaluate both swap directions for every pair, yielding 120 directional interventions per model.

We consider two downstream conditions. In the \emph{judgment-fixed} condition, the explicit judgment is held at \texttt{BLOCK} in both baseline and patched trajectories before the common \texttt{Final action:} query. In the second condition, the explicit judgment is determined from the model's ALLOW-versus-BLOCK readout before and after the state swap and then supplied to the same action query. Because the two swap directions have opposite semantics, we report donor-aligned effects: positive judgment alignment indicates movement of the judgment readout toward the donor condition, and positive action alignment indicates movement of action preference toward the donor condition.

\begin{table*}[t]
\centering
\small
\setlength{\tabcolsep}{3pt}

\resizebox{0.72\textwidth}{!}{%
\begin{tabular}{lccc}
\toprule
Model & Judgment alignment & Action, judgment fixed & Action, label unchanged \\
\midrule
Mistral L24 &
$+0.06042$ [$+0.05052,+0.06979$] &
$+0.00004$ [$-0.00063,+0.00071$] &
$-0.00070$ [$-0.00147,+0.00007$] \\
Qwen L32 &
$+0.01563$ [$-0.00417,+0.03594$] &
$-0.00027$ [$-0.00099,+0.00047$] &
$-0.00039$ [$-0.00122,+0.00041$] \\
Gemma L32 &
$+0.41042$ [$+0.31771,+0.50729$] &
$+0.00051$ [$-0.00090,+0.00196$] &
$+0.00013$ [$-0.00149,+0.00173$] \\
\bottomrule
\end{tabular}%
}

\caption{\textbf{Naturally occurring judgment-state variation transfers weakly to downstream action.}
Matched counterfactual states are swapped at the judgment position. Judgment alignment measures movement toward the donor judgment condition; action columns report donor-aligned downstream effects when the explicit judgment is fixed or when it remains unchanged after the swap. Values are means with pair-cluster bootstrap 95\% confidence intervals over 60 held-out matched pairs.}
\label{tab:app-natural-state}
\end{table*}

Natural state replacement measurably shifts the judgment readout in Mistral and Gemma while leaving downstream action preference essentially unchanged when the explicit judgment is fixed (Table~\ref{tab:app-natural-state}). Qwen shows similarly negligible action effects at L32, while its judgment-side shift is small at this layer. The same near-zero action pattern holds when the explicit label is allowed to vary but remains unchanged.

Mistral additionally shows a sharp change when the state swap crosses the explicit judgment boundary. The 10 directional interventions that change the ALLOW/BLOCK label produce a donor-aligned action effect of $+0.13459$ [95\% CI $+0.11724,+0.15182$], whereas interventions that leave the label unchanged remain near zero ($-0.00070$). Boundary crossings are rare at the corresponding Qwen configuration (1/120) and absent at Gemma L32. Thus the naturally occurring judgment state need not behave as a persistent downstream action-control variable even when it changes the judgment readout; when an intervention does alter the explicit judgment supplied to subsequent computation, substantially stronger coupling can emerge.

\paragraph{Aggregation and uncertainty.}
The statistical unit is the matched scenario pair. Each bootstrap replicate samples 60 pair identities with replacement and retains both directional interventions associated with every sampled pair. We report percentile 95\% confidence intervals from 10,000 pair-cluster bootstrap replicates. Label-switch subsets are used descriptively to localize where the downstream action response is concentrated.

\section{Learned Recombination Within the Judgment-Control Subspace}
\label{sec:app-recombination}

\subsection{Per-case learned recombination}

The learned-recombination analysis compares four directions: the original judgment-derived direction, the learned within-judgment-subspace controller \(\cJA\), the projected train-derived action-controller component \(P_{\UJ}\cA\), and the full action-native controller \(\cA\).
The Mistral analysis shown here uses Split A and evaluates its held-out \kbd subset (\(N=35\)) at L16/L20/L24/L32.

\begin{figure*}[t]
    \centering
    \includegraphics[width=\textwidth]{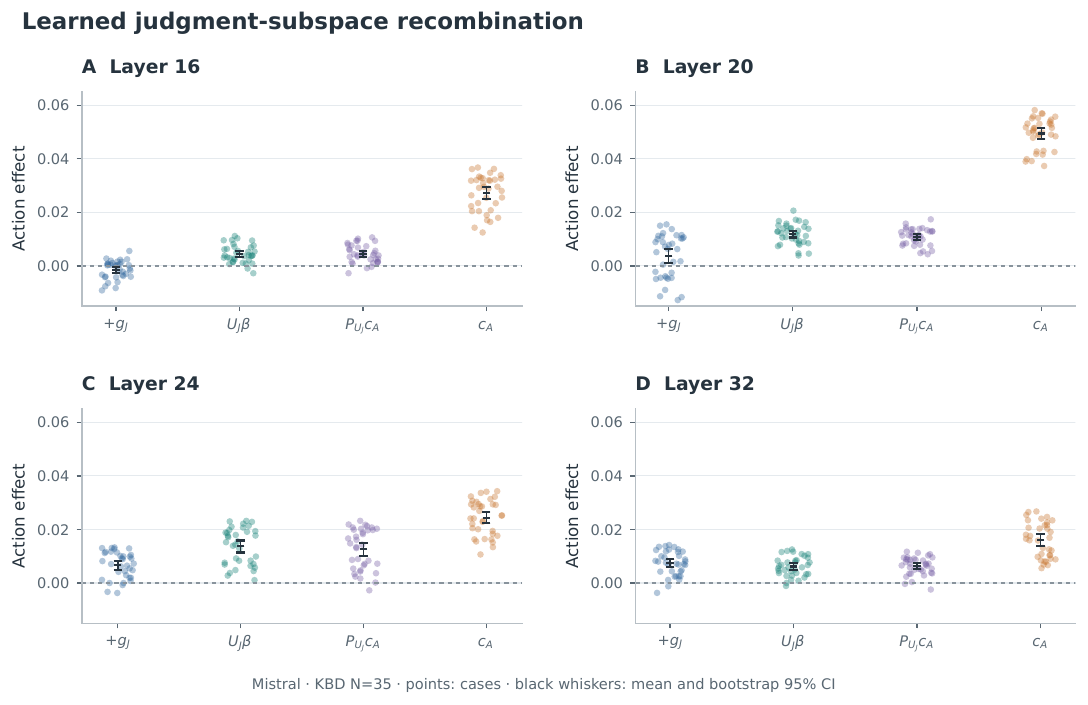}
    \caption{\textbf{Per-case learned judgment-subspace recombination.}
    At Mistral middle layers, the learned within-subspace action readout and the projected action-controller component have similar held-out distributions, are typically stronger than the single judgment direction, and remain weaker than the full action controller.}
    \label{fig:app-a11}
\end{figure*}

For Mistral, \(\cJA\) yields 0.00446, 0.01188, 0.01369, and 0.00617 at L16/L20/L24/L32, while \(P_{\UJ}\cA\) yields 0.00452, 0.01084, 0.01264, and 0.00637.
The corresponding full action-controller effects are 0.02717, 0.04948, 0.02445, and 0.01606.
The learned and projected components share the same train-family judgment subspace and action-gradient information; their agreement therefore shows recovery of the same train-derived shared component rather than an independent source of evidence.

\section{Judgment Sign-Flip Correctness Control}
\label{sec:app-sign-control}

The sign-control experiments verify the semantics of the judgment gradient independently of its transfer to action.
By definition,
\[
\gJraw=\nabla[\log p(\texttt{ALLOW})-\log p(\texttt{BLOCK})],
\]
so \(+\gJraw\) is ALLOW-oriented and \(-\gJraw=\gJsafe\) is BLOCK-oriented on blocked-valid cases.
We evaluate both signs at seven approximately matched relative depths (20--80\%) in all three architectures: Qwen L13/L19/L26/L32/L38/L45/L51, Gemma L12/L19/L25/L31/L37/L43/L50, and Mistral L8/L12/L16/L20/L24/L28/L32.
All three models use the complete Split A held-out \texttt{blocked\_valid} population (\(N=60\)) for this matched-depth control.

\paragraph{Judgment endpoint.}
Across all three architectures and all tested depths, the mean judgment effect has the expected orientation: \(+g_J\) increases the raw \texttt{ALLOW}$-$\texttt{BLOCK} margin and \(-g_J\) decreases it.
For Qwen and Gemma, case-wise sign consistency reaches 100\% from roughly 40\% relative depth onward; for Mistral, the two signs produce opposite judgment effects on all 60 held-out cases at every tested depth.
The following main-analysis layers provide representative numerical checks (the matched-density sweep uses the layer grid above):
\begin{center}
\small
\begin{tabular}{llrrrr}
\toprule
Model & Layer & \(\Delta m_J(+g_J)\) & \(\Delta m_J(-g_J)\) & \(\Delta m_A(+g_J)\) & \(\Delta m_A(-g_J)\) \\
\midrule
Qwen & L32 &  0.9734 & -0.8958 &  0.00345 & -0.00280 \\
Qwen & L48 &  4.5974 & -4.3604 &  0.00068 & -0.00112 \\
Gemma & L24 & 1.9063 & -2.3396 & -0.00194 &  0.00114 \\
\bottomrule
\end{tabular}
\end{center}
Here \(m_A\) is the raw unsafe-target action margin, so the safe-oriented effect reported elsewhere is \(-\Delta m_A\).
These controls confirm that \(\gJsafe\) is a correctly oriented and depth-robust controller of explicit safety judgment.

\paragraph{Action endpoint.}
The transferred action response does not show the same global antisymmetry.
For Qwen, the mean raw action effects of \(+g_J\) and \(-g_J\) have opposite signs at all seven tested depths, but the case-wise opposite-sign fraction is only about 25--55\%.
For Gemma, mean action responses are oppositely signed from L12 through L31, whereas from L37 onward both signs produce negative mean raw action effects, yielding clear late-depth non-antisymmetry.
Mistral shows the same distinction particularly clearly: the judgment endpoint is perfectly sign-reversed case-wise at every tested depth, whereas action-side opposite-sign fractions range from 16.7\% to 66.7\% and vary non-monotonically with depth.

\begin{table*}[t]
\centering
\small
\begin{tabular}{rrrrrrr}
\toprule
Layer & \(\Delta m_J(+g_J)\) & \(\Delta m_J(-g_J)\) & J opposite sign & \(\Delta m_A(+g_J)\) & \(\Delta m_A(-g_J)\) & A opposite sign \\
\midrule
L8  & 0.192708 & -0.189583 & 60/60 (100.0\%) &  0.000525 &  0.000224 & 18/60 (30.0\%) \\
L12 & 0.320313 & -0.304688 & 60/60 (100.0\%) &  0.001153 & -0.000565 & 20/60 (33.3\%) \\
L16 & 1.701042 & -1.440625 & 60/60 (100.0\%) &  0.001430 & -0.001112 & 20/60 (33.3\%) \\
L20 & 4.333333 & -3.796354 & 60/60 (100.0\%) & -0.001667 &  0.001096 & 40/60 (66.7\%) \\
L24 & 5.619271 & -6.526563 & 60/60 (100.0\%) & -0.003565 & -0.000309 & 20/60 (33.3\%) \\
L28 & 6.449479 & -7.043815 & 60/60 (100.0\%) & -0.003391 & -0.003046 & 10/60 (16.7\%) \\
L32 & 6.293490 & -7.063041 & 60/60 (100.0\%) & -0.003834 &  0.001809 & 32/60 (53.3\%) \\
\bottomrule
\end{tabular}
\caption{\textbf{Matched-depth Mistral sign control on the complete Split A held-out population.}
Judgment effects reverse sign on every held-out case at every tested depth, while action-side antisymmetry is partial and depth dependent. Action columns report the raw unsafe-target margin effect; lower values are safer. Excluding exact-zero action responses leaves the same qualitative pattern (16.9--67.8\% opposite-sign across depth).}
\label{tab:app-mistral-sign-control}
\end{table*}

Thus sign reversal is a strong correctness check on the native judgment endpoint but not a general law of cross-interface action response.
The matched-depth controls separate two claims that should not be conflated: the judgment intervention has stable ALLOW/BLOCK semantics on its native endpoint, whereas its action-side sensitivity is weaker, architecture dependent, and only locally antisymmetric in parts of the network.

\section{Relative Judgment-to-Action Efficacy Across Intervention Strength}
\label{sec:app-action-ratio}

\begin{figure*}[t]
    \centering
    \includegraphics[width=0.96\textwidth]{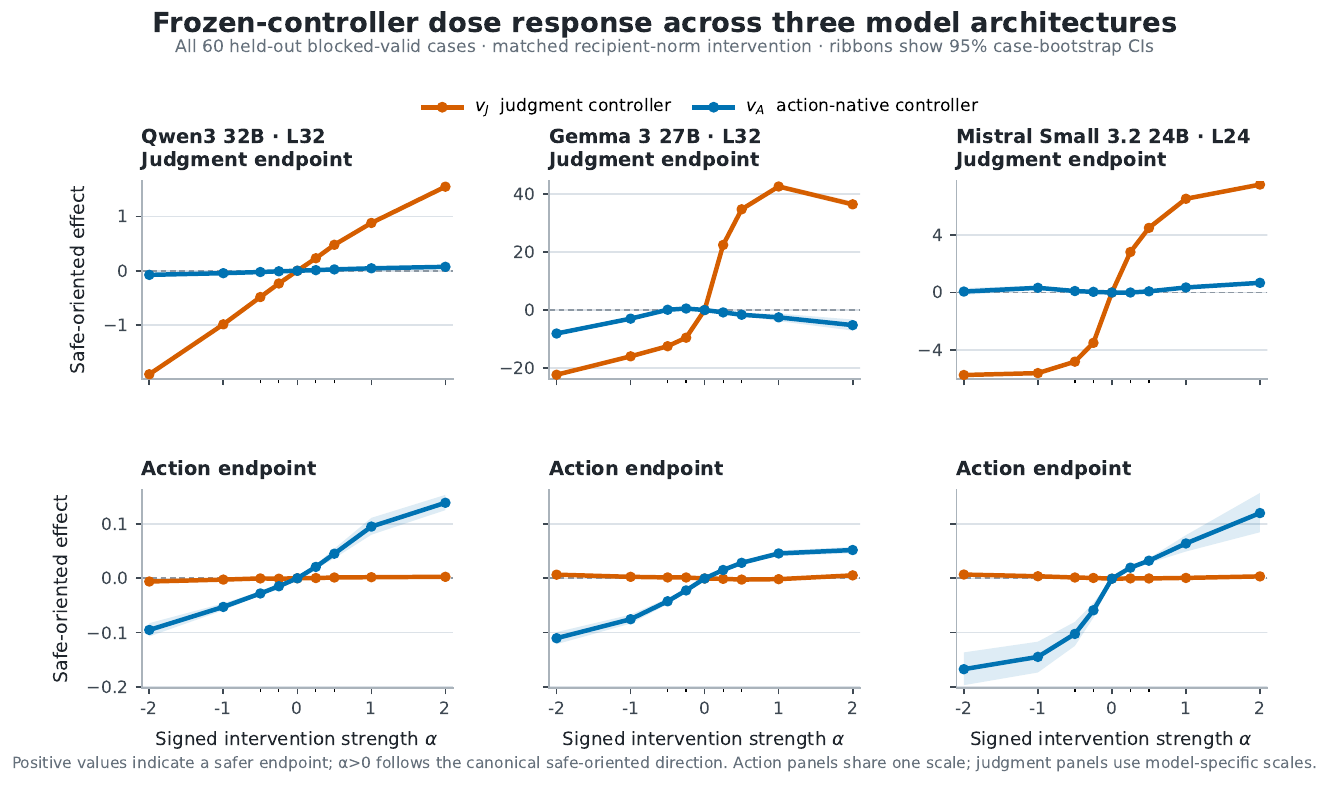}
    \caption{\textbf{Judgment and action controllers exhibit asymmetric causal response profiles across intervention strength.}
    Frozen train-derived controllers are evaluated on all 60 Split A held-out \texttt{blocked\_valid} scenarios at Qwen L32, Gemma L32, and Mistral L24.
    Signed \(\alpha\) scales the recipient-norm-matched intervention; positive values follow the canonical safe-oriented direction.
    The judgment controller \(v_J\) strongly controls its native judgment endpoint but remains weak on action, whereas the action-native controller \(v_A\) strongly controls action and can also affect judgment with architecture-dependent sign and magnitude.
    Ribbons denote case-level bootstrap 95\% confidence intervals.}
    \label{fig:app-controller-dose-response}
\end{figure*}

Figure~\ref{fig:app-controller-dose-response} shows the full signed dose-response curves that support the main-text summary.
At \(\alpha=1\), the judgment controller \(v_J\) produces safe-oriented judgment effects of \(0.8813\), \(42.6643\), and \(6.523\) for Qwen, Gemma, and Mistral, respectively, while its action-side effects are only \(0.00199\), \(-0.00165\), and \(0.00066\). On the same action endpoint and at the same matched strength, the action-native controller \(v_A\) yields \(0.09507\), \(0.04567\), and \(0.06405\).

On the action endpoint, we summarize the dose-response separation by
\begin{equation}
R(\alpha)
=
\frac{\left|\SafeEff(\alpha v_J)\right|}
{\left|\SafeEff(\alpha v_A)\right|},
\qquad \alpha\neq 0,
\label{eq:relative-action-efficacy}
\end{equation}
which compares judgment-derived and action-native efficacy at the same signed intervention strength.
Across Qwen, Gemma, and Mistral, \(R(\alpha)\) remains small throughout the tested range, with values below \(0.11\) in all three architectures.

\begin{figure*}[t]
    \centering
    \includegraphics[width=0.88\textwidth]{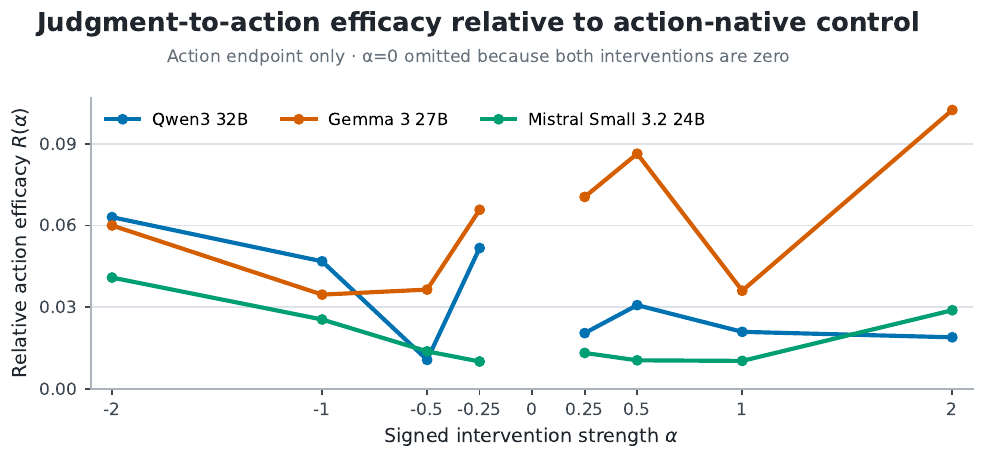}
    \caption{\textbf{Judgment-to-action efficacy remains a small fraction of action-native control.}
    We report \(R(\alpha)=|\SafeEff(\alpha v_J)|/|\SafeEff(\alpha v_A)|\) on the action endpoint for each nonzero signed intervention strength.
    Across Qwen, Gemma, and Mistral, the judgment-derived controller remains far less effective on action than the action-native controller over the full tested range.}
    \label{fig:action-transfer-ratio}
\end{figure*}

\section{Blocked-Valid Population and KBD Stratification}
\label{sec:app-full-population}

Figure~\ref{fig:causal-depth} reports the complete 60-scenario held-out \texttt{blocked\_valid} population for each model under Split A.
Here we additionally stratify the same held-out population by the baseline \kbd phenotype, defined before intervention.

\paragraph{Qwen3-32B.}
The Split A held-out population contains 16 \kbd and 44 non-\kbd scenarios.
The action-native controller remains substantially stronger in both groups and at both tested layers:
\begin{center}
\footnotesize
\setlength{\tabcolsep}{2pt}
\begin{tabular}{rlccc}
\toprule
Layer & Group & \(\SafeEff(-g_J)\) [95\% CI] & \(\SafeEff(c^A)\) [95\% CI] & Paired gap [95\% CI] \\
\midrule
L32 & All 60 & 0.00280 [0.00115, 0.00449] & 0.09523 [0.08012, 0.11127] & 0.09243 [0.07702, 0.10969] \\
    & \kbd 16 & 0.00173 [-0.00095, 0.00459] & 0.13147 [0.09960, 0.16331] & 0.12974 [0.09771, 0.16296] \\
    & non-\kbd 44 & 0.00319 [0.00112, 0.00520] & 0.08205 [0.06691, 0.09908] & 0.07887 [0.06337, 0.09676] \\
\addlinespace
L48 & All 60 & 0.00112 [-0.00024, 0.00248] & 0.05895 [0.04599, 0.07303] & 0.05783 [0.04523, 0.07233] \\
    & \kbd 16 & 0.00136 [-0.00024, 0.00297] & 0.07875 [0.05250, 0.10648] & 0.07739 [0.05201, 0.10551] \\
    & non-\kbd 44 & 0.00103 [-0.00072, 0.00277] & 0.05175 [0.03763, 0.06791] & 0.05072 [0.03697, 0.06672] \\
\bottomrule
\end{tabular}
\end{center}

\paragraph{Gemma3-27B.}
All 60 Split A held-out blocked-valid scenarios satisfy the \kbd definition, so the full-population and \kbd evaluations coincide and there is no within-model non-\kbd subgroup.
\begin{center}
\small
\begin{tabular}{rccc}
\toprule
Layer & \(\SafeEff(-g_J)\) [95\% CI] & \(\SafeEff(c^A)\) [95\% CI] & Paired gap [95\% CI] \\
\midrule
L32 & -0.00206 [-0.00379,-0.00043] & 0.04677 [0.04132,0.05224] & 0.04883 [0.04388,0.05395] \\
L48 &  0.00052 [-0.00230, 0.00334] & 0.01667 [0.01052,0.02293] & 0.01615 [0.00989,0.02246] \\
\bottomrule
\end{tabular}
\end{center}

\paragraph{Mistral-Small-3.2-24B.}
The Split A held-out population contains 35 \kbd and 25 non-\kbd scenarios.
Representative layers are shown below; the action-native controller remains stronger in both subgroups throughout the L8--L32 sweep.
\begin{center}
\small
\begin{tabular}{llrrrr}
\toprule
Layer & Held-out group & \(N\) & \(\SafeEff(-g_J)\) & \(\SafeEff(c^A)\) & \(c^A-(-g_J)\) \\
\midrule
L16 & All blocked-valid & 60 &  0.0011 & 0.0267 & 0.0256 \\
    & \kbd             & 35 &  0.0021 & 0.0267 & 0.0246 \\
    & non-\kbd         & 25 & -0.0002 & 0.0267 & 0.0269 \\
\addlinespace
L20 & All blocked-valid & 60 & -0.0011 & 0.0505 & 0.0516 \\
    & \kbd             & 35 & -0.0009 & 0.0497 & 0.0507 \\
    & non-\kbd         & 25 & -0.0013 & 0.0517 & 0.0530 \\
\addlinespace
L24 & All blocked-valid & 60 &  0.0003 & 0.0633 & 0.0630 \\
    & \kbd             & 35 &  0.0007 & 0.0241 & 0.0234 \\
    & non-\kbd         & 25 & -0.0003 & 0.1181 & 0.1183 \\
\addlinespace
L32 & All blocked-valid & 60 & -0.0018 & 0.0158 & 0.0176 \\
    & \kbd             & 35 & -0.0021 & 0.0163 & 0.0184 \\
    & non-\kbd         & 25 & -0.0013 & 0.0151 & 0.0164 \\
\bottomrule
\end{tabular}
\end{center}

Thus the controller ordering observed in the complete blocked-valid population is also present separately in \kbd and non-\kbd cases for Qwen and Mistral.
For Mistral, the same ordering is additionally preserved when the 60 held-out cases are divided into three equal-sized strata by baseline action margin.

\section{Split-Half Stability of Control Subspaces}
\label{sec:app-subspace-stability}

The low overlap between judgment- and action-control subspaces could in principle arise if each estimated space were itself unstable under finite-sample variation.
All headline stability analyses therefore use Split B of the causal set.
For each model, tested layer, target tool, and rank \(k\in\{1,2,4,8,16\}\), train-family scenarios are split into two halves while preserving tool and family structure as closely as possible.
We independently construct \(U^{J,(1)}_{\ell,k},U^{J,(2)}_{\ell,k}\) and \(U^{A,(1)}_{\ell,k},U^{A,(2)}_{\ell,k}\), and compute
\[
S_{JJ}=S(U^{J,(1)},U^{J,(2)}),\qquad
S_{AA}=S(U^{A,(1)},U^{A,(2)}),\qquad
S_{JA}=S(U^{J,(1)},U^{A,(2)}),
\]
using the overlap score from Equation~\ref{eq:subspace-overlap}.
Qwen and Gemma use 30 random split-half repetitions per configuration; the Split B Mistral analysis uses 50 repetitions.
At rank 16, each half contains roughly 20 train gradients per target tool.

At rank 16, averaged over target tools and tested layers, the results are:
\begin{center}
\begin{tabular}{lrrrr}
\toprule
Model & \(S_{JJ}\) & \(S_{AA}\) & \(S_{JA}\) & Random expectation \\
\midrule
Qwen3-32B & 52.94\% & 54.56\% & 1.86\% & 0.3125\% \\
Gemma3-27B & 47.33\% & 47.42\% & 4.70\% & 0.2976\% \\
Mistral-Small-3.2-24B & 49.81\% & 42.14\% & 1.71\% & 0.3125\% \\
\bottomrule
\end{tabular}
\end{center}

For Mistral, the Split B rank-16 layerwise \(J\)--\(A\) overlap decreases from 3.16\% at L8 to below 1\% at L24--L32, while the corresponding within-computation \(J\)--\(J\) and \(A\)--\(A\) reproducibility remains substantially larger.
Across all three models, the within-computation split-half overlap is therefore much higher than the cross-computation overlap.
The low judgment--action overlap is not explained simply by finite-sample instability of the estimated subspaces.
The spaces are not completely orthogonal: Gemma, in particular, retains more cross-computation sharing, consistent with the larger late-layer overlap reported in the main text.

\section{Action-Controller-to-Judgment Reverse Transfer}
\label{sec:app-action-to-judgment}

We additionally test the reverse intervention \(\cA\rightarrow J\).
This supplementary analysis evaluates the held-out \kbd subsets associated with the corresponding controller analyses (\(N=30\) for Qwen, \(N=60\) for Gemma, and \(N=35\) for Mistral), rather than the complete 60-scenario Split A population used in Figure~\ref{fig:causal-depth}.
The action controller \(\cA\) is constructed only from train-family blocked-valid scenarios, conditioned on target tool, frozen, and injected into the held-out \kbd judgment interface.
The judgment endpoint is the original margin
\[
\jmargin=\log p(\texttt{ALLOW})-\log p(\texttt{BLOCK}),
\]
and we report the safe-oriented judgment effect \(-\Delta\jmargin\), so positive values indicate movement toward \texttt{BLOCK}.
The intervention modifies the final judgment-prompt token at a single residual position and is norm matched to the corresponding held-out case's allowed-valid minus blocked-valid judgment-state displacement.
The analysis uses the same layers and controller hyperparameters as the corresponding forward-transfer experiments.

\begin{table*}[t]
\centering
\small
\begin{tabular}{llrrr}
\toprule
Model & Layer & Safe-oriented \(\cA\rightarrow J\) & 95\% CI & Safe-positive \\
\midrule
Qwen3-32B & L32 & -0.0167 & [-0.0458,\;0.0125] & 5/30 \\
Qwen3-32B & L48 &  0.0500 & [ 0.0125,\;0.0875] & 13/30 \\
\addlinespace
Gemma3-27B & L32 & -4.3667 & [-4.7250,\;-4.0333] & 0/60 \\
Gemma3-27B & L48 & -8.5625 & [-8.8451,\;-8.2740] & 0/60 \\
\addlinespace
Mistral-Small-3.2-24B & L8  &  0.0277 & [ 0.0196,\;0.0357] & 27/35 \\
Mistral-Small-3.2-24B & L12 &  0.0357 & [ 0.0268,\;0.0446] & 27/35 \\
Mistral-Small-3.2-24B & L16 &  0.0670 & [ 0.0571,\;0.0768] & 34/35 \\
Mistral-Small-3.2-24B & L20 & -0.2134 & [-0.3938,\;-0.0286] & 15/35 \\
Mistral-Small-3.2-24B & L24 &  0.2094 & [ 0.1500,\;0.2701] & 27/35 \\
Mistral-Small-3.2-24B & L28 &  0.6147 & [ 0.5888,\;0.6411] & 35/35 \\
Mistral-Small-3.2-24B & L32 &  0.3040 & [ 0.2522,\;0.3567] & 35/35 \\
\bottomrule
\end{tabular}
\caption{\textbf{Reverse transfer from action-native control to explicit judgment is architecture- and depth-dependent.}
Positive values shift the explicit judgment toward \texttt{BLOCK}; negative values shift it toward \texttt{ALLOW}.}
\label{tab:app-action-to-judgment}
\end{table*}

The reverse-transfer pattern is not a symmetric double dissociation.
Qwen is closest to a weak-transfer regime: L32 is statistically compatible with zero, while L48 has a small positive mean effect.
Gemma shows a qualitatively different pattern: the action controller produces a large and completely consistent ALLOW-oriented judgment shift at both tested layers.
Mistral changes sign with depth, with positive effects at L8--L16, a negative effect at L20, and positive effects again at L24--L32.
Thus action-native controllers can be null, safety-oriented, or oppositely oriented on the judgment interface.
The result therefore qualifies control separation as an interface-dependent causal relationship rather than causal independence between judgment and action computations.

\end{document}

%% file: math_commands.tex
\usepackage{amsmath,amsfonts,bm}

\def\eqref#1{equation~\ref{#1}}

\def\1{\bm{1}}

\DeclareMathAlphabet{\mathsfit}{\encodingdefault}{\sfdefault}{m}{sl}
\SetMathAlphabet{\mathsfit}{bold}{\encodingdefault}{\sfdefault}{bx}{n}

